\documentclass{elsart}
\usepackage{epsfig}
\usepackage{amssymb}
\journal{Physics Letters B}
\def\v#1{\mbox{\boldmath$#1$}}

\newcommand{\mc}{\multicolumn}
\newcommand{\lsim}{\mathrel{\mathop{\kern 0pt \rlap
  {\raise.2ex\hbox{$<$}}}
  \lower.9ex\hbox{\kern-.190em $\sim$}}}
\newcommand{\gsim}{\mathrel{\mathop{\kern 0pt \rlap
  {\raise.2ex\hbox{$>$}}}
  \lower.9ex\hbox{\kern-.190em $\sim$}}}

\begin{document}
\begin{frontmatter}
\title{On the effect of the
$\Delta(1232)$ in hypernuclear non--mesonic weak decay: a microscopic approach}

\author{E. Bauer$^1$ and G. Garbarino$^2$}

\address{$^1$Departamento de F\'{\i}sica, Universidad Nacional de La Plata and
IFLP, CONICET C. C. 67, 1900 La Plata, Argentina}

\address{$^2$Dipartimento di Fisica, Universit\`a di Torino, I--10125 Torino, Italy}

\date{\today}

\begin{abstract}
The non--mesonic weak decay of $\Lambda$--hypernuclei is studied within a
microscopic diagrammatic approach which includes,
for the first time, the effect or the $\Delta$--baryon resonance. We adopt a nuclear matter
formalism extended to finite nuclei via the local density approximation, a one--meson
exchange weak transition potential, a Bonn nucleon--nucleon strong potential
and a $\Delta N\to NN$ strong potential based on the Landau--Migdal theory.
Ground state correlations and final state interactions (FSI), at second order in the
baryon--baryon strong interaction, are introduced on the same footing
for all the isospin channels of one-- and two--nucleon induced decays.
Weak decay rates and single and double--coincidence nucleon spectra are predicted for
$^{12}_\Lambda$C and compared with recent KEK and FINUDA data.
The $\Delta(1232)$ introduces new FSI--induced decay mechanisms which lead to an
improvement when comparing the obtained nucleon spectra with data,
while it turns out to have a negligible effect on the decay rates.
Discrepancies with experiment remain only for emission spectra involving protons, but
are mostly restricted to double--nucleon correlations in the non--back--to--back kinematics.
\end{abstract}
\end{frontmatter}

The study of hypernuclear non--mesonic weak decay is of fundamental importance, since it
provides primary means of exploring the four--baryon, strangeness changing, weak
interactions \cite{AG02}.
The determination of these interactions, which also have a relevant impact
in the physics of dense stars \cite{Sch10}, requires the solution of complex many--body
problems together with a big amount of correlated information from a very systematic and
coordinated series of measurements.
The non--mesonic decay width, $\Gamma_{\rm NM} = \Gamma_1 + \Gamma_2$, is built up
from one-- (1N) and two--nucleon
induced (2N) decays, $\Gamma_1 = \Gamma_n + \Gamma_p$ and
$\Gamma_2 = \Gamma_{nn} + \Gamma_{np} + \Gamma_{pp}$, where the isospin components
are given by
$\Gamma_N = \Gamma (\Lambda N \to nN)$ and $\Gamma_{NN'} = \Gamma (\Lambda NN' \to nNN')$,
with $N$, $N' =n$ or $p$.

After several decades during which the experimental information was scarce, due to the
difficulties inherent in the design of the experiments, in recent years or so some very
interesting new measurements were
carried out. In particular, we refer to the recent single--nucleon and nucleon--coincidence
experiments carried out at KEK \cite{KEK0,KEK1,KEK2,KEK3} and FINUDA \cite{FINUDA1,FINUDA2}.
These advances were accompanied by the advent of elaborated theoretical models
(some of which included final state interactions and ground state correlations
effects on the same ground) and allowed us to reach a reasonable agreement between data and
predictions for the non--mesonic weak decay rates and asymmetry parameters
(for a recent review see \cite{Rev12}).

However, discrepancies between theory and experiment
are still present for the emission spectra involving protons \cite{BGPR,BG11}.
Concerning the theory--experiment disagreement, a further comment is in order.
The extraction of decay rates (for instance, the $\Gamma_n/\Gamma_p$ and
$\Gamma_2/\Gamma_{\rm NM}$ ratios) and asymmetries from data is done by a theoretical
analysis of the nucleon emission spectra (the real observables),
which are affected by nucleon final state interactions.
A disagreement between theory and experiment for the spectra should thus be reflected
by discrepancies among the decay rates. This does not occur at present: probably,
the explanation of this outcome is hidden beyond the big experimental error bars
and/or in some inadequacy of the theoretical frameworks.
From the theoretical viewpoint,
it is still unclear the role played by the $\Delta$--baryon resonance,
the first excited state of the nucleon, and
by possible violations of the $\Delta I=1/2$ isospin rule in the non--mesonic decay
\cite{Rev12}. Further theoretical work is thus needed.
New experiments will be carried out at J--PARC, GSI and FAIR (HypHI
Collaboration), while new analyses are expected also from FINUDA.

In this Letter we further extend the diagrammatic approach developed
in~\cite{BG11,Ba07,Ba10b} to include the $\Delta(1232)$ baryon.
The electromagnetic properties of this resonance have been extensively studied in recent
years. The relatively small mass difference between the nucleon and the $\Delta$
together with the strong coupling of the resonance with the $\pi N$ channel
implies for the $\Delta$ a relevant role in strong
interaction physics; this was clearly demonstrated, for instance, in studies of
few--nucleon forces, nuclear phenomena, heavy--ion collisions and neutron stars.
Here we study the relevance of this resonance in non--mesonic weak hypernuclear decay. A
nuclear matter formalism is adopted and results for the decay rates and especially
the single and double--coincidence
nucleon spectra are reported for $^{12}_\Lambda$C within the local density approximation.
Ground state correlations (GSC)
and nucleon final state interactions (FSI) contributions
are introduced at second order in the nucleon--nucleon and
$\Delta N\to NN$ strong interactions for the whole set of 1N and 2N isospin decay channels.
Being fully quantum--mechanical, the present approach
turned out to produce more reliable results \cite{BG11} for FSI than
those based on the (semi--classical) nucleon rescattering given by
intranuclear cascade (INC) models \cite{Ra97,Ga03}.

A clear advantage of the microscopic approach is that
nucleon FSI and GSC can be included on the same footing
in the calculation of decay widths and nucleon emission spectra.
This approach thus incorporates a consistency level which cannot be achieved in INC calculations.
While GSC give rise to 2N decay amplitudes which mainly contribute to the
non--mesonic decay rate, FSI are important in the evaluation of the nucleon spectra.
In addition, as demonstrated in \cite{BG11}, the microscopic model naturally contains
quantum interference terms (QIT) among different weak decay amplitudes
which are of fundamental importance.

The disadvantage of the microscopic approach is that it is much more intricate
than the INC. Any observable is evaluated by
considering a certain set of many--body Goldstone diagrams.
Various diagrams must be tested until one finds the most relevant ones at any given
level of approximation. Previous calculations are used as a guidance to improve the
predictions by adding new contributions. In~\cite{BG11}, we were able to reproduce rather well the
spectra for neutron emission, while our predictions largely overestimated the spectra involving
protons. Also, the obtained results improved the INC ones.

Only nucleon degrees of freedom were considered in~\cite{BG11} to build up
the many--body Goldstone diagrams for the $\Lambda$ self--energy.
No other paper has ever included any nucleon resonance.
We therefore decided to improve the set of diagrams with the inclusion of the
$\Delta(1232)$ baryon. We expect the relevance of the $\Delta$ in non--mesonic weak decay
to be smaller than in other nuclear physics processes like pion or electron scattering off
nuclei. Indeed, the $Q$--value of the non--mesonic decay,
$\simeq m_\Lambda-m_n= 177$~MeV, is not enough to produce a $\Delta$ in the final state.
The $\Delta$ plays a role only when it is off--shell, producing both GSC and FSI
contributions. The resonance has isospin $3/2$ and thus appears in four charge states:
$\Delta^-$, $\Delta^0$, $\Delta^+$ and $\Delta^{++}$.

\begin{figure}[t]
\begin{center}
\mbox{\epsfig{file=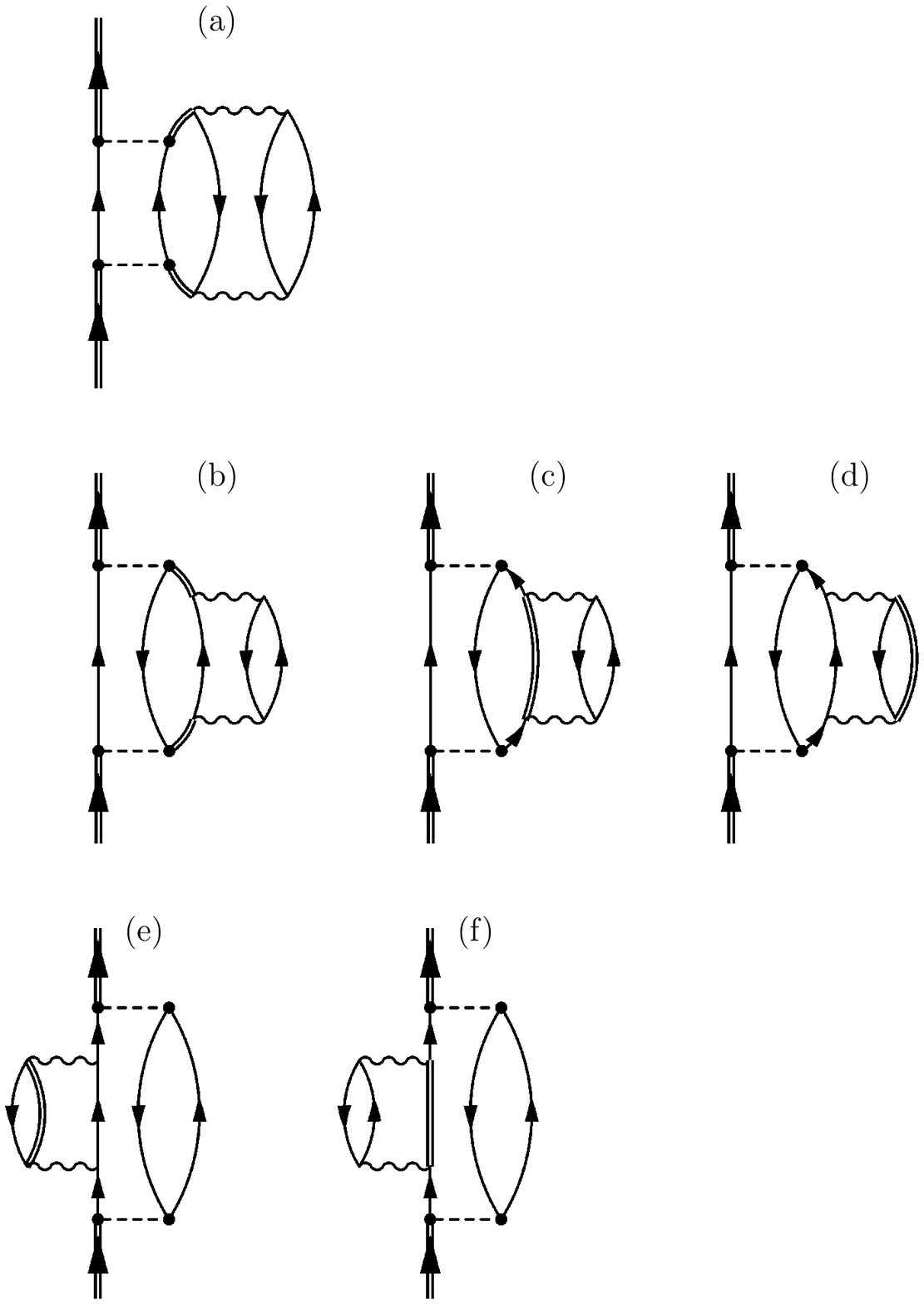,width=.63\textwidth}}
\caption{The set of Goldstone diagrams for the $\Lambda$ self--energy considered in this
work. The dashed and wavy lines stand for the potentials $V^{\Lambda N \to NN}$ and
$V^{\Delta N\to NN}$, respectively.}
\label{fig1}
\end{center}
\end{figure}

We thus have to choose the set of $\Lambda $ self--energy diagrams containing the $\Delta(1232)$ to
be evaluated in addition to the nucleonic contributions already discussed in~\cite{BG11}
and given in Fig.1 of that reference.
Due to energy--momentum conservation, the nucleon resonance cannot contribute to 1N decay
amplitudes at zeroth order in the strong interaction.
Higher order GSC diagrams also contributing to 1N decay amplitudes and incorporating the
$\Delta$ can be neglected since,
on the basis of former studies on hypernuclei~\cite{Ba10b} and electron scattering~\cite{Ba98},
they are expected to provide much smaller contributions than the two-- and
three--nucleon emission diagrams given in Fig.~\ref{fig1} which we consider here.

Since, as explained, the nucleon resonance cannot appear in the final state,
any cut crossing a $\Delta$ provides a vanishing contribution to decay rates and
spectra. Diagrams (a) and (b) contain two $\Delta$'s and
give rise to a 2N decay and a FSI--induced decay, respectively.
From the numerical analysis it turns out that these two diagrams have small effects.
For instance, the contribution to the two--nucleon decay rate $\Gamma_2$ in $^{12}_\Lambda$C
of diagram (a) is 0.002 in units of the free $\Lambda$ decay rate $\Gamma_\Lambda^{\rm free}$,
while $\Gamma_2=0.36\, \Gamma_\Lambda^{\rm free}$. The contribution of diagrams (a) and (b)
to the nucleon emission spectra is of the same order of magnitude.
Diagrams (b) to (f) corresponds to FSI--induced decays, each one of them admitting only
a single final state. Both the (a) and (b) contributions admit a $3p2h$ final state (three--nucleon
emission). For all the remainder terms one has $2p1h$ final states (two--nucleon emission).
Diagrams (c) to (f) are QIT: each one of them is the product of two different
$\Lambda N\to nN$ transition amplitudes. From the calculation it turns out that
the (c)--(f) terms provide negative contributions; instead,
diagrams (a) and (b) give positive contributions (although small) by construction.

From this discussions it turns out that the hypernuclear weak decay rates remain unchanged
with respect to the results given in Table 1 of~\cite{BG11}, which agree rather well
with the recent KEK \cite{KEK3} and FINUDA \cite{FINUDA1,FINUDA2} data.
In the present Letter we do not give explicit expressions for the
self--energies of Fig.~\ref{fig1}. However, some general features
have to be discussed to clarify the particular isospin structure
of these $\Delta$ contributions.
Let us thus first write the total number of
nucleons and nucleon pairs emitted in the non--mesonic decay
as follows~\cite{Ba07}:
\begin{eqnarray}
\label{nn1f}
N_{n} &= & 2 \bar{\Gamma}_{n} +  \bar{\Gamma}_{p} +
3 \bar{\Gamma}_{nn} + 2 \bar{\Gamma}_{np} + \bar{\Gamma}_{pp}
+ \sum_{i, \, f} N_{f \, (n)}
\bar{\Gamma}_{i, f}~, 
\\
\label{np1f}
N_{p} & = & \bar{\Gamma}_{p} + \bar{\Gamma}_{np} + 2
\bar{\Gamma}_{pp} +
\sum_{i, \, f} N_{f \, (p)} \, \bar{\Gamma}_{i, f}~,
\\
\label{nnn1f}
N_{nn} & = & \bar{\Gamma}_{n} + 3 \bar{\Gamma}_{nn}
+ \bar{\Gamma}_{np}+
\sum_{i, \, f} N_{f \, (nn)} \, \bar{\Gamma}_{i, f}~,
\\
\label{nnp1f}
N_{np} & = & \bar{\Gamma}_{p} + 2 \bar{\Gamma}_{np}
+ 2 \bar{\Gamma}_{pp} +
\sum_{i, \, f} N_{f \, (np)} \, \bar{\Gamma}_{i, f}~,
\\
\label{npp1f}
N_{pp} & = & \bar{\Gamma}_{pp} +
\sum_{i, \, f} N_{f \, (pp)} \, \bar{\Gamma}_{i, f}~,
\end{eqnarray}
where a normalization per non--mesonic decay is used
($\bar{\Gamma} \equiv \Gamma/\Gamma_{\rm NM}$).
Single and double coincidence nucleon spectra are obtained by constraining the
evaluation of each $\bar \Gamma$ to certain intervals in energy, opening angle, etc.
The $\bar \Gamma_N$'s ($\bar \Gamma_{NN'}$'s) are the 1N (2N) decay rates, while
the remaining terms containing the functions $\bar{\Gamma}_{i, f}$
represent FSI Goldstone diagrams.
The index $i$ in $\bar{\Gamma}_{i, f}$ is used to label the various FSI Goldstone
diagrams included in the present calculation. We remind the reader that,
apart from the $\Delta$ contributions
of Fig.~\ref{fig1}, also the ones of Fig.~1 in~\cite{BG11} are considered here.
The index $f$ in $\bar{\Gamma}_{i, f}$
instead denotes the final physical state of the Goldstone diagram and in the
present case can take the values $f=nN$ (cut on $2p1h$ states) and $nNN'$ (cut on $3p2h$
states). Finally, $N_{f \, (N)}$ ($N_{f \, (NN')}$) is the number of
nucleons of the type $N$ (of $NN'$ pairs) contained in the multinucleon state $f$.
Concerning Eqs.~(\ref{nn1f})--(\ref{npp1f}), we also note that diagram (a) of Fig.~\ref{fig1}
provides a (small) contribution to the decay rates $\bar \Gamma_{NN'}$, while the FSI--induced
diagrams (b) to (f) contribute to the functions $\bar{\Gamma}_{i, f}$.

Before discussing the numerical results we give some detail on the adopted
weak and strong potentials. The weak transition potential $V^{\Lambda N\to NN}$
contains the exchange of the full set of
mesons of the pseudoscalar ($\pi$, $\eta$, $K$) and vector octets ($\rho$, $\omega$,
$K^*$), with strong coupling constants and cut--off parameters deduced from
the Nijmegen soft--core interaction NSC97f~\cite{St99}.
For the nucleon--nucleon interaction $V^{NN\to NN}$ we adopt the Bonn potential
(with the exchange of $\pi$, $\rho$, $\sigma$ and $\omega$ mesons) \cite{Ma87}.
Finally, we have modeled the $\Delta N\to NN$ strong potential in terms of
attractive $\pi$ and $\rho$ meson exchange potentials complemented by a repulsive
Landau--Migdal term driven by the $g'_{\Delta N}$ constant:
\begin{eqnarray}
\label{deltaint}
V^{\Delta N\to NN}(q) & = & \frac{f_{\pi NN} f_{\pi \Delta N}}{m^{2}_{\pi}}
\,
\Gamma_{\pi NN}(q) \Gamma_{\pi \Delta N}(q) [g'_{\Delta N}
\v{\sigma} \cdot \v{S}
+ \frac{q^{2}}{q^{2}+m^{2}_{\pi}} \,
\v{\sigma} \cdot \hat{\v{q}} \v{S} \cdot \hat{\v{q}}  \nonumber \\
&&+ \frac{\Gamma_{\rho NN}(q) \Gamma_{\rho \Delta N}(q)}{\Gamma_{\pi NN}(q)
\Gamma_{\pi \Delta N}(q)}
\, C_{\rho} \,
\frac{q^{2}}{q^{2}+m^{2}_{\rho}} \,
(\v{\sigma} \times \hat{\v{q}}) \cdot (\v{S} \times \hat{\v{q}})]  \,
\v{\tau} \cdot \v{T}\, ,
\end{eqnarray}
where with $\Gamma_{\pi X N}(q)=(\Lambda^{2}_{\pi X N}-m^{2}_{\pi})/(\Lambda^{2}_{\pi X N}+q^{2})$
(an analogous expression holds for $\Gamma_{\rho X N}$), with $X=N$ or $\Delta$, we denote
the hadronic form factors, $\Lambda_{\pi NN}=\Lambda_{\pi \Delta N}=1300$~MeV/c,
$\Lambda_{\rho NN}=1400$~MeV/c and $\Lambda_{\rho \Delta N}=1700$~MeV/c being the
correponding cut--offs. Moreover, $g'_{\Delta N}=0.4$ and for the hadronic coupling
constants we have: $f^{2}_{\pi NN}/4 \pi = 0.081$, $f_{\pi \Delta N} = 2 f_{\pi NN}$ and
$C_{\rho}= f_{\rho NN} f_{\rho \Delta N}/(f_{\pi NN} f_{\pi \Delta N})=2.18$.

\begin{figure}[h]
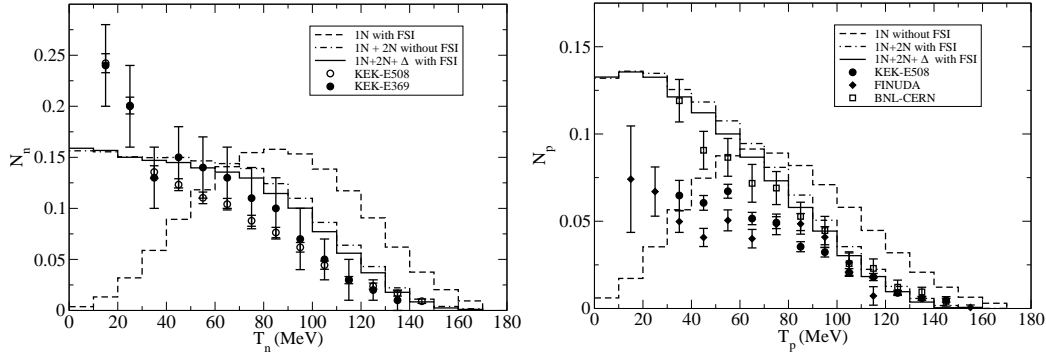

\begin{center}
\mbox{\epsfig{file=NnTn.eps,width=.49\textwidth}}
\mbox{\epsfig{file=NpTp.eps,width=.49\textwidth}}
\caption{Neutron and proton kinetic energy spectra for $^{12}_\Lambda$C
non--mesonic
weak decay. The dashed (dot--dashed, continuous) lines are normalized
per 1N decay (per non--mesonic decay). Experimental data are from
KEK--E369~\protect\cite{Kim03}, KEK--E508~\protect\cite{Ok04},
FINUDA~\protect\cite{FINUDA1} and BNL--CERN~\protect\cite{Mo74}.}
\label{NpTp}
\end{center}
\end{figure}
\begin{figure}[h]
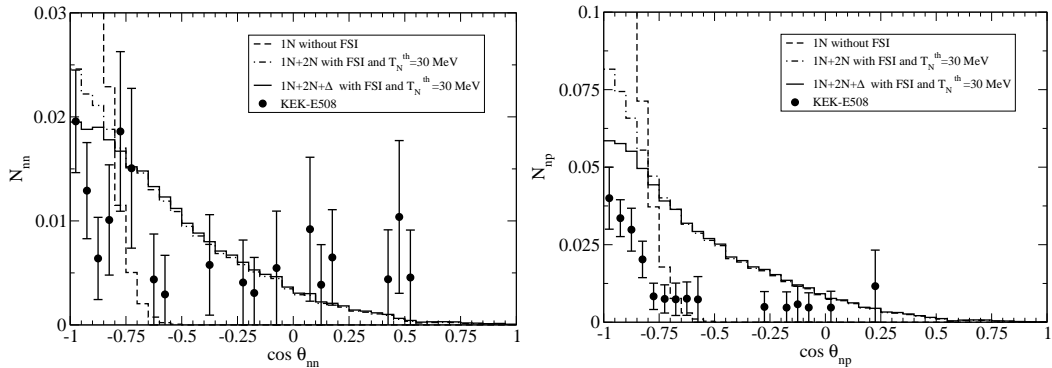

\begin{center}
\mbox{\epsfig{file=Nnnc.eps,width=.49\textwidth}}
\mbox{\epsfig{file=Nnpc.eps,width=.497\textwidth}}
\caption{Opening angle distribution of $nn$ and $np$ pairs.
Normalization is as in Fig.~\protect\ref{NpTp}. Data are from
KEK--E508~\protect\cite{Kim06}.}
\label{Nnpc}
\end{center}
\end{figure}
\begin{figure}[h]
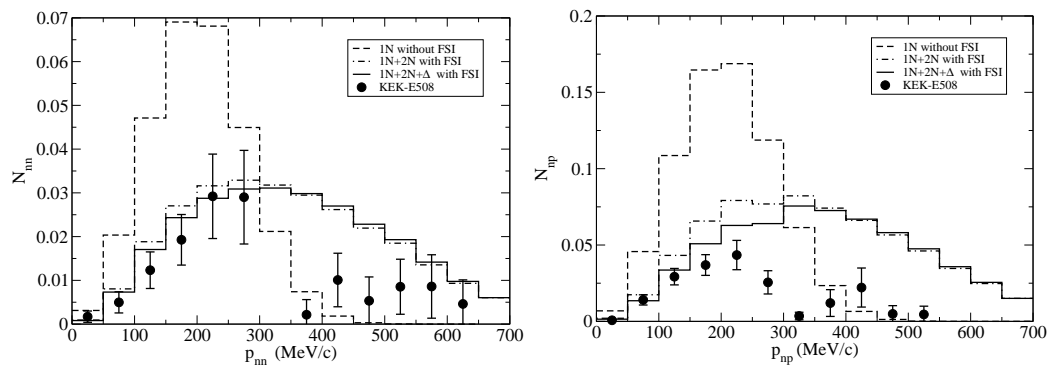

\begin{center}
\mbox{\epsfig{file=Nnnp12.eps,width=.49\textwidth}}
\mbox{\epsfig{file=Nnpp12.eps,width=.49\textwidth}}
\caption{Momentum correlation spectra of $nn$ and $np$ pairs,
with $p_{NN'}\equiv |\vec{p}_{N}+\vec{p}_{N'}|$.
Normalization is as in Fig.~\protect\ref{NpTp}. Data are
from KEK--E508~\protect\cite{Kim09}.}
\label{Nnpp12}
\end{center}
\end{figure}
\begin{figure}[h]
\begin{center}
\mbox{\epsfig{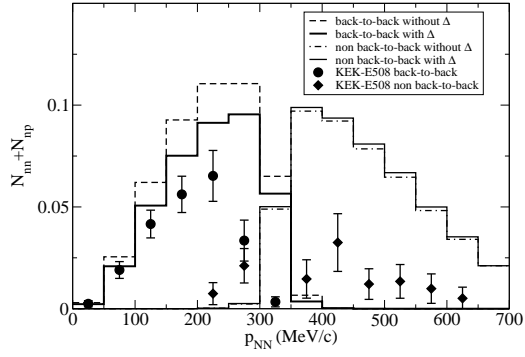}}
\caption{Momentum correlation spectra for the sum of the
$nn$ and $np$ pair numbers for the back--to--back
($\cos \theta_{NN}<-0.7$) and non back--to--back
kinematics ($\cos \theta_{NN}>-0.7$). The continuous lines refer to the
full calculation, while the dashed and dot--dashed lines are obtained
without considering the $\Delta$. Normalization is per non--mesonic weak decay.
Data are from KEK--E508~\protect\cite{Kim10}.}
\label{Nnnnpp12}
\end{center}
\end{figure}

We turn now to our main concern in this Letter: the study of the nucleon emission spectra.
In Fig.~\ref{NpTp} we show the neutron and proton
kinetic energy spectra for the non--mesonic decay
of $^{12}_\Lambda$C. The dashed curves are the distributions
of the 1N decay nucleons (normalized per 1N decay):
as expected, they show a maximum at half of the $Q$--value
for $^{12}_\Lambda$C non--mesonic decay and a bell--type shape
due to the nucleon Fermi motion and the $\Lambda$ momentum distribution
in the hypernucleus. The inclusion of 2N and FSI--induced decay processes
provides the results given by the dot--dashed lines (normalized per non--mesonic decay)
and leads to a reduction of the nucleon average energy,
thus filling the low--energy part of the spectrum and emptying the high--energy
region. This outcome has been explained in detail
in~\cite{BG11}\footnote{Note that the present ``1N$+$2N with FSI'' results
of Figs.~\ref{NpTp}--\ref{Nnnnpp12} slightly differ from the ones obtained in~\cite{BG11}.
This is due to a more refined numerical analysis of those $2p1h$ nucleonic contributions
of Fig.1 in~\cite{BG11} which are obtained by regularizing divergent integrals by
the Cauchy principal value method.}. The ``1N$+$2N with FSI'' results
reproduce fairly well the KEK neutron spectra, while
a rather strong overestimation is found of the
KEK--E508 and FINUDA proton distributions. Our ``1N$+$2N with FSI'' proton spectrum is instead
closer to the old BNL--KEK data.
The final results, containing the $\Delta$ contributions, are given by continuous lines
(again, they are normalized per non--mesonic decay).
Starting from $T_N\simeq 30$ MeV, these spectra are lower than the purely nucleonic ones by a
non--negligible amount, thus improving the comparison with data. In particular,
the theoretical proton spectrum now agrees well with the old BNL--KEK data, but still overestimates
the recent measurements. However, one must note that
a certain dispersion is clearly visible in Fig.~\ref{NpTp} among the three
experimental proton spectra (the discrepancies concerns not only the magnitude of the spectrum but
also its shape). Possibly, new proton data could clarify these results.

The opening angle distributions of $nn$ and $np$ pairs
are reported in Fig.~\ref{Nnpc}.
To adhere to the KEK data, the predictions of the calculations including FSI are
obtained for a 30 MeV nucleon kinetic energy threshold $T^{\rm th}_N$.
The distributions arising from the 1N decay (dashed curves)
are strongly peaked at $\theta_{NN'}=180^\circ$ (back--to--back kinematics).
The ``1N$+$2N with FSI'' results (dot--dashed
lines) show that the QIT have a crucial effect: they considerably reduce the back--to--back
contribution and strongly populates the non back--to--back region.
The final ``1N$+$2N$+\Delta$ with FSI'' spectra are given by continuous lines.
The diagrams which incorporate the nucleon resonance turn out to reduce the
spectra in the back--to--back region
and improve the comparison with data. We note that
the ``1N$+$2N with FSI'' and ``1N$+$2N$+\Delta$ with FSI'' results
turn out to be very sensitive to the value adopted for $T^{\rm th}_N$.
The agreement with KEK--E508 data is rather good for the $nn$ spectrum,
while for $np$ pairs a significant overestimation is still present. The latter result is
compatible with the overestimation of the proton spectrum obtained in the same experiment
(see Fig.~\ref{NpTp}).

In Fig.~\ref{Nnpp12} we give the two--nucleon momentum correlation spectra,
i.e., the ${nn}$ and ${np}$ distributions as a function of the
momentum sum $p_{NN'}\equiv|\vec{p}_{N}+\vec{p}_{N'}|$
of two of the outgoing nucleons. The dashed lines correspond to the 1N decay;
the dot--dashed curves refer to the results with 1N, 2N
and FSI included; finally, the continuous curves show the full,
``1N$+$2N$+\Delta$ with FSI'' predictions.
Both the ``1N$+$2N with FSI'' and ``1N$+$2N$+\Delta$ with FSI'' calculations
are performed by considering a nucleon kinetic energy threshold
$T^{\rm th}_N=30$ MeV, as in the data also shown in the figures.
The maximum at $p_{NN'}\simeq 200$ MeV/c of the 1N distributions displaces to larges
$p_{NN'}$ values do to 2N and FSI--induced decays, which indeed produce less back--to--back
peaked events.
As noted in~\cite{Kim09}, the minimum in both the ${nn}$ and ${np}$
KEK--E508 distributions is an effect of the low statistics
and detection efficiency for events with $p_{NN'}\gsim$ 350 MeV/c
(the KEK detector geometry being optimized for back--to--back coincidence events,
i.e., for small values of $p_{NN'}$).
Indeed, such dip structure has not been found in our calculation,
which overestimates the data for large correlation momenta
(especially for the ${np}$ spectrum, consistently with the spectra discussed so far).
Moreover, according to our calculation, a double--maximum structure as suggested by the data
could possibly be explain only with very big GSC and/or FSI, but this
would spoil the agreement in the momentum correlation spectra found here
at small $p_{NN'}$ as well as the agreement found for the previous spectra.
Finally, note that the $\Delta$ has the effect of improving the comparison with
data for small values of $p_{NN'}$, especially in the ${np}$ case.

The ``1N$+$2N with FSI'' and ``1N$+$2N$+\Delta$ with FSI''
distributions of Fig.~\ref{Nnpp12} at low momentum sum
(say below 400 MeV/c) are mainly due to 1N decays
(which are strongly back--to--back correlated), while for higher momenta
the contribution of  2N and FSI--induced decays
is dominant (and produces less back--to--back correlated pairs).
This behavior is confirmed by the momentum correlation of the sum
$N_{nn}+N_{np}$ shown in Fig.~\ref{Nnnnpp12}
for the opening angle regions with $\cos \theta_{NN'}<-0.7$ (back--to--back region)
and $\cos \theta_{NN'}>-0.7$ (non back--to--back region).
Again, one notes an improvement in the comparison with data in the back--to--back
region thanks to the $\Delta$ contributions.

Finally, in Table~\ref{number-ratio} we show results for the
$N_{nn}/N_{np}$ ratio obtained with a nucleon kinetic energy threshold of 30 MeV
for the back--to--back ($\cos \theta_{NN'}<-0.7$) and the non--back--to--back regions
($\cos \theta_{NN'}>-0.7$).
An improvement, although limited, in the comparison with KEK--E508 data is noted for the
calculation which includes the $\Delta$. Our final results for both angular regions
are compatible with data for $N_{nn}/N_{np}$ within $\simeq 1.8\, \sigma$. The underestimation
of this ratio originates from the overestimation of the $N_{np}$ spectrum
and confirms a systematic overestimation of the proton emission reported by KEK--E508.
\begin{table}[h]
\begin{center}
\caption{The $N_{nn}/N_{np}$ ratio is given for the back--to--back
($\cos\, \theta_{NN'}< -0.7$) and non back--to--back regions
($\cos\, \theta_{NN'}> -0.7$) in the case of $T^{\rm th}_N=30$ MeV.
Data are from KEK--E508~\protect\cite{Kim06}.}
\label{number-ratio}
\begin{tabular}{l c c| c} \hline\hline
\mc {1}{c}{Angular region} &
\mc {1}{c}{Without $\Delta$} &
\mc {1}{c|}{With $\Delta$} &
\mc {1}{c}{KEK--E508} \\ \hline
$\cos\, \theta_{NN'}< -0.7$ & 0.33 & 0.37 & $0.60\pm 0.12$ \\
$\cos\, \theta_{NN'}> -0.7$ & 0.39 & 0.42 & $1.38\pm 0.53$ \\
\hline\hline
\end{tabular}
\end{center}
\end{table}


Summarizing, a microscopic approach including for the first time many--body terms
introduced by the nucleonic resonance $\Delta(1232)$ is used to evaluate the
nucleon emission spectra in non--mesonic weak decay of hypernuclei.
Such a scheme provides a fully quantum--mechanical description in
which a unified treatment of complex effects such as GSC and nucleon FSI is considered.
As we have seen in a previous Letter \cite{BG11},
in this scheme QIT play a key role. This is confirmed by the
present calculation: among the Goldstone diagrams incorporating the $\Delta$, the
relevant ones are QIT and turn out to produce a sensitive reduction of the spectra.
On the contrary, the new diagrams have a negligible effect in the calculation of the
non--mesonic decay rates.
Although an improvement is achieved thanks to the new weak decay channels,
especially concerning coincidence spectra in the back--to--back kinematics,
discrepancies with experiment still remain for proton emission.
Further work is in order to understand the origin of the disagreement.
Forthcoming coincidence experiments at J--PARC on
$^{12}_\Lambda$C \cite{JparcE18} and four--body hypernuclei \cite{JparcE22}
will allow a measurement of the nucleon spectra with improved accuracy.
From the theoretical viewpoint, new studies should consider a possible violation
of the $\Delta I = 1/2$ rule on the isospin change in the non--mesonic decay.
For the future we plan to start a systematic investigation
of rare non--mesonic weak decays of $\Lambda \Lambda$--hypernuclei such as
$\Lambda \Lambda \to  \Lambda n$,
$\Lambda \Lambda \to  \Sigma^- p$,
$\Lambda \Lambda \to  \Sigma^0 n$ (which implies a strangeness variation $\Delta S=1$) and
$\Lambda \Lambda \to  n n$ ($\Delta S=2$).
A reliable calculation of the rates for
these $\Lambda$--induced $\Lambda$ decay reactions is missing, no experimental
evidence of such processes is available at present but they could be observed in the
future.	


\end{document}